\documentclass[conference]{IEEEtran}
\IEEEoverridecommandlockouts
\usepackage{cite}
\usepackage{amsmath,amssymb,amsfonts}
\usepackage{algorithmic}
\usepackage{graphicx}
\usepackage{textcomp}
\usepackage{xcolor}
\usepackage{multirow}
\usepackage{listings}
\usepackage{rotating}
\def\BibTeX{{\rm B\kern-.05em{\sc i\kern-.025em b}\kern-.08em
    T\kern-.1667em\lower.7ex\hbox{E}\kern-.125emX}}
\begin{document}
\title{Augmenting Machine Learning with Information Retrieval to Recommend Real Cloned Code Methods for Code Completion}
\author{\IEEEauthorblockN{Muhammad Hammad}
\IEEEauthorblockA{\textit{Eindhoven University of Technology} \\
Netherlands\\
m.hammad@tue.nl}
\and
\IEEEauthorblockN{\"{O}nder Babur}
\IEEEauthorblockA{\textit{Eindhoven University of Technology} \\
Netherlands\\
o.babur@tue.nl}
\and
\IEEEauthorblockN{Hamid Abdul Basit}
\IEEEauthorblockA{\textit{Prince Sultan University} \\
Saudi Arabia \\
hbasit@psu.edu.sa}
% \and
% \IEEEauthorblockN{Mark van den Brand}
% \IEEEauthorblockA{\textit{Eindhoven University of Technology} \\
% Netherlands \\
% m.g.j.v.d.brand@tue.nl}
}

\maketitle
%%%%%%%%%%%%%%%%%
%%%%%%%%%I would like to mention following papers, which are quite relevant to our current paper, which will benefit co-authors to understand how other researchers belonging to different domains make the output from language model in a readable form.....our approach is also similar to them.
% Generating sentences by editing prototypes
% https://arxiv.org/abs/1709.08878
% Search engine guided neural machine translation
% https://www.aaai.org/ocs/index.php/AAAI/AAAI18/paper/download/17282/16060
% Two are better than one: An ensemble of retrieval-and generation-based dialog systems
% https://arxiv.org/abs/1610.07149
% Domain-specific image captioning
% https://www.aclweb.org/anthology/W14-1602.pdf
% A retrieve-and-edit framework for predicting structured outputs
% https://papers.nips.cc/paper/8209-a-retrieve-and-edit-framework-for-predicting-structured-outputs.pdf

\begin{abstract}
Software developers frequently reuse source code from repositories as it saves development time and effort. Code clones accumulated in these repositories hence represent often repeated functionalities and are candidates for reuse in an exploratory or rapid development. In previous work, we introduced DeepClone, a deep neural network model trained by fine tuning GPT-2 model over the BigCloneBench dataset to predict code clone methods. The probabilistic nature of DeepClone output generation can lead to syntax and logic errors that requires manual editing of the output for final reuse. In this paper, we propose a novel approach of applying an information retrieval (IR) technique on top of DeepClone output to recommend real clone methods closely matching the predicted output. We have quantitatively evaluated our strategy, showing that the proposed approach significantly improves the quality of recommendation.
\end{abstract}

% similar approaches based on language models. \cite{sim1998archetypal} 

\section{Introduction}
%Help developers develop code very fast---previous work we develop DeepClone but problem in neural language technique
% such as speech recognition \cite{creutz2007morph}, machine translation \cite{jean2014using}, comment generation\cite{hu2018deep}, fault detection \cite{ray2016naturalness}, code completion \cite{mou2015end}\cite{nguyen2013statistical}, code clone detection \cite{svajlenko2016machine,white2016deep}, code search \cite{gu2018deep} and code summarization \cite{iyer2016summarizing}mason2014domain guu2018generating gu2018search
Software developers need effective code search and reuse capability for rapid or exploratory development \cite{sadowski2015developers}, as writing source code from scratch is an expensive activity. Often, programming of well-defined features amounts to a simple look-up in one's own or others' code in repositories. With the increasing volume of available source code repositories and online resources, it gets more probable to find useful code snippets \cite{gabel2010study}. Nevertheless, for identifying the relevant parts of the code for reuse\cite{sim1998archetypal}, developers turn to ad-hoc code reuse\cite{juergens2009code} with manual searching and selective reading of the source code. It is an expensive and error-prone activity without effective support mechanisms like code snippet search, code predictions, code auto-completion and code generation, to assist them in writing code quickly and correctly. Language modeling is among the most popular methods to realize these features \cite{mou2015end,nguyen2013statistical,gu2018deep}.

%The source code, however, possesses certain differences (e.g.~ non-linear structure and large volume) from natural language text, making it difficult for the developers to read it conveniently. 

% Following the observation on the reuse potential of code clones,
In previous work we proposed DeepClone, a deep neural network model trained by fine tuning GPT-2 over the BigCloneBench code clone dataset, for the purpose of clone method predictions. Despite having promising results from various evaluations, DeepClone had a shortcoming: the generated code snippets can contain syntax and logic errors due to the probabilistic nature of the language model, and the specific neural language generation technique applied (nucleus sampling\cite{holtzman2019curious}). This eventually requires manual modification making its reuse burdensome (see Table \ref{tab:example1} and Table \ref{tab:example2} for an example). 

%However, the desired output might be a variation of another, previously observed sample \cite{song2016two,hashimoto2018retrieve}. 

% . neural language generation techniques \cite{li2017adversarial,shao2017generating}. However, the desired output might be a variation of another, previously observed sample \cite{guu2018generating,gu2018search,song2016two,hashimoto2018retrieve,mason2014domain}.

%%%%Limitation of text generation
This problem is not specific to DeepClone, but rather an inherent problem of language models. In natural language generation, it is a well-known challenge to generate well-formed outputs \cite{li2017adversarial,shao2017generating}. While recently there are significant advancements in neural language generation techniques, they still cannot match the quality of human authored content (e.g.~programs or texts) \cite{fan2018hierarchical}. They further possess certain problems at their core, notably, standard likelihood training and decoding leads to dull and repetitive outputs \cite{holtzman2019curious}, and more training data and advanced sampling techniques does not seem to solve this issue entirely \cite{radford2019language}. Token-level probabilities predicted by the language models also remain relatively poor \cite{welleck2019neural}. However, the desired output might be a variation of another, previously observed sample \cite{song2016two,hashimoto2018retrieve}.

% An extension would involve displaying the most similar cloned methods (as is) from the dataset to the user.  after replacing meta tokens such as  \big \langle num\_val\big\rangle~and  \big \langle str\_val\big\rangle~with real values, and removing meta tokens such as \big \langle soc\big\rangle~and  \big \langle eoc\big\rangle~(see our previous paper\cite{hammad2020deepclone} for details). 

Language models (in our context) are fundamentally probabilistic models, which can generate multiple possible sequences of output (in our case clone methods) based on user context. The space of possible clone methods that could be generated grows exponentially with the length of the clone methods. By having $V$ tokens in the vocabulary, there can be $V^N$ possible clone methods of length $N$ that could be generated. DeepClone model also has a similar problem as it generates clone methods that differ from the real clone methods, which can lead to various syntax and logic errors. For instance, in Table \ref{tab:example2}, "destDir" identifier has been declared in the DeepClone output, but it has not been used anywhere. A fully trained language model can learn patterns in the code such as opening and closing brackets, but it cannot completely learn the logical flow of the code. This motivates our work here, where we seek to build a system that can recommend real clone methods on the basis of user context. Here, a real clone method is taken from a real project, contains the code of some particular functionality "as is", and has been manually validated by the curators of BigCloneBench. Our approach combines the DeepClone model output and information retrieval (IR) techniques to recommend real clone methods. 

% Examples would include an identifier being declared at the start, but not being used later on.which will compile without syntax errors,
%%%%%%%%%%%%Benefit of code clones%%%%%%%%%%
Recommending code clones has various benefits. Code clones are useful for exploratory development, where the rapid development of a feature is required and the remedial unification of newly generated clone is not clearly justified \cite{kapser2008cloning}. Also, cloned code is expected to be more stable and poses less risk than new development. Hence, we believe that clone methods can be considered a useful component for neural code generation, as they can be used to capture the common code practices of developers, which can be offered as code prediction and completion to the developer. 

In this work, we improve the re-usability aspect of code clones by recommending real clone methods using IR techniques to remove noise in the clone methods predicted by neural code generation. We believe that our approach can help in improving the quality of clone prediction on the basis of user context. In this paper, we have made the following contributions:

%Hence, it can be argued that developers need to reuse code for desired software features in a way that supports opportunistic programming for increased productivity
\begin{enumerate}
 \item We present a novel approach for recommending real code clone methods by augmenting the DeepClone model output using an IR technique (TF-IDF) for retrieving most similar clone methods from a search corpus.
 \item We have quantitatively evaluated our approach in terms of accuracy and effectiveness by calculating various metrices. The overall results show that the proposed approach significantly improves the quality of recommendation from DeepClone. 
\end{enumerate}

%These LMs used for code prediction typically target predicting the next token based on a window of subsequence of input. In our previous study, we observed the potential to exploit code clones to enhance prediction capabilities. Code clones are repeated patterns in the code, which are usually created with the copy-and-paste practice mentioned above.  According to Roy and Cordy \cite{roy2007survey}, around 5 to 50\% of the code in software applications can be contained in clones, which can be of several granularity levels such as line, method, file, and directory. These clones can be considered as a useful component for a language model, as they can be used to capture the common code practices of developers, which then can be used for code predictions and completions to the new developer. Kasper and Godfrey \cite{kapser2008cloning} observed that code clones can have positive effects on the software development in certain circumstances. One of the positive use cases is exploratory development, where the rapid development of a feature is required and the remedial unification of newly generated clone is not clearly justified. Also, a piece of cloned code is expected to be more stable and poses less risk than new development. Hence, it can be argued that developers need to reuse code for desired software features in a way that supports opportunistic programming for increased productivity. 

\section{Methodology}
\begin{figure*}
  \centering
  \includegraphics[width=12cm,height=9cm]{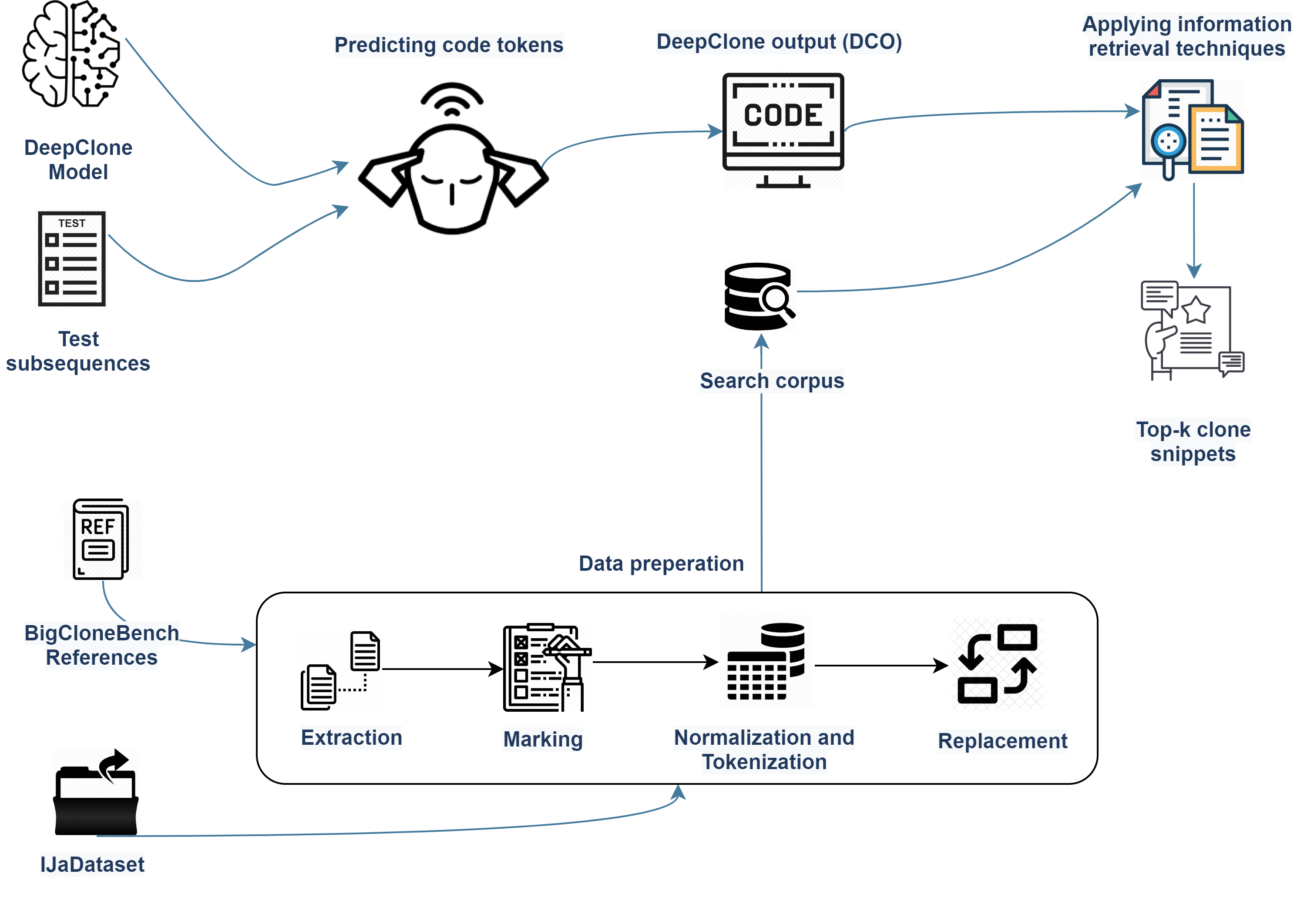}
  \caption{Methodology of generating real code clones\label{fig:methodology}}
\end{figure*}

We propose a methodology for recommending real clone methods on the basis of given code context by applying IR technique over the DeepClone output. Our starting point is the DeepClone model previously trained on the datasets BigCloneBench and IJaDataset. We have also developed a search corpus comprising of real clone methods from those datasets. Initially, independent of the search corpus, we attempt to predict clone methods consisting of subsequent token sequences starting from the start-of-clone tag \big \langle soc\big\rangle~ until the end-of-clone tag \big \langle eoc\big\rangle~(\cite{hammad2020deepclone}). Then, we apply an IR technique to retrieve real clone methods from the search corpus, which are most similar to the initially generated DeepClone prediction. Figure ~\ref{fig:methodology} displays a pictorial representation of our methodology to generate real clone methods (without showing the training steps of DeepClone model, which are shown in our previous paper\cite{hammad2020deepclone}). In their raw form, we obtain DeepClone output, ground truth, and top-10 samples from the current methodology in an un-formatted style, where code tokens of a clone method are separated only by space characters. To make these outputs readable, we have formatted the code by using online tool \footnote{https://www.tutorial spoint.com//online\_java\_formatter.htm} along with little manual editing. Table ~\ref{tab:example1} and Table ~\ref{tab:example2} contain the formatted output. We describe the details of our methodology in the following subsections. %, which we plan to automate in future. 

\subsection{Building the Search Corpus}
\label{sec:bigclonebench}%svajlenko2015evaluating svajlenko2016bigcloneeval
We have built our search corpus from BigCloneBench and IJaDataset\cite{svajlenko2014towards,svajlenko2016bigcloneeval}, which we have also previously used to train DeepClone. BigCloneBench consists of over 8 million manually validated clone method pairs in IJaDataset 2.0 \cite{IJaDataset:2020:Online}- a large Java repository of 2.3 million source files (365 MLOC) from 25,000 open-source projects. BigCloneBench contains clones with both syntactic and semantic similarities, along with the references of starting and ending lines of method clones existing in the code repository. In forming this benchmark, methods that potentially implement a given common functionality were identified using pattern based heuristics. These methods were manually tagged as true or false positives of the target functionality by multiple judges. All true positives of a functionality were grouped as a clone class, where a clone class of size $n$ contains $\frac{n(n-1)}{2}$ clone pairs. Currently, BigCloneBench contains clones corresponding to 43 distinct functionalities. Further details can be found in the relevant publications \cite{svajlenko2014towards,svajlenko2016bigcloneeval}.
% BigCloneBench has been primarily developed to measure and compare the recall of clone detection tools \cite{svajlenko2014towards,svajlenko2015evaluating,svajlenko2016bigcloneeval,sajnani2016sourcerercc}. However, it can also be used for other clone and software studies \cite{svajlenko2016bigcloneeval}. Li et al. \cite{li2017cclearner}, for instance, have developed a DNN-based clone detector, CCLearner, using BigCloneBench. 

We have performed several pre-processing steps to build our search corpus, which is similar to what we have followed in our previous work \cite{hammad2020deepclone}. First, we have extracted the details of a total of 14,922 true positive clone methods, in which 11,920 are distinct clone methods (\textit{Extraction}). Next, we have traced them in IJaDataset files, by following their references from the BigCloneBench dataset, and put them in our search corpus list by placing meta tokens \big \langle soc\big\rangle~at the start, and \big \langle eoc\big\rangle~at the end of each clone method (\textit{Marking}). These meta tokens are also part of the DeepClone output, so inserting them in the search corpus clone method list helps in making a fair comparison. Afterwards, we have normalized each clone method code by removing whitespaces, extra lines, comments, as well as tokenizing (\textit{Normalization and Tokenization}) by adapting Javalang\footnote{https://github.com/c2nes/javalang} Python library, which contains a lexer and parser for Java 8 programming language. Finally, for each clone method, we have replaced integer, float, binary, and hexadecimal constant values with the \big \langle num\_val\big \rangle~meta-token (\textit{Replacement}). Similarly, we have replaced string and character values with \big \langle str\_val\big \rangle. Again, this is just to ensure to have fair comparison as DeepClone output is also in this normalized format.

%Neural Text generation https://transformer.huggingface.co/doc/gpt2-large ficler2017controlling
\subsection{Clone Method Prediction by DeepClone}
 Several neural language generation methods can be used to predict token subsequences from clone methods based on the user context, such as beam search \cite{vijayakumar2018diverse}, sampling with temperature \cite{ackley1985learning}, top-k sampling \cite{fan2018hierarchical} and nucleus sampling \cite{holtzman2019curious}. Each generation method has a specific decoding strategy to shape the probability distribution of language model, such as assigning higher probability to higher quality text. We have used the nucleus sampling method in DeepClone model as it outperforms other methods and is commonly considered the best strategy for generating large amounts of high quality text, comparable to human written text\cite{holtzman2019curious}. By having GPT-2 fine-tuned for DeepClone model, along with nucleus sampling (threshold 0.95), we expect to generate a coherent set of code tokens for clone method predictions. Holtzman et al. \cite{holtzman2019curious} have also achieved coherent text generation results with similar settings.

We have performed small scale (100 context queries) experiment to perform next token subsequence prediction by choosing different subsequence sizes such as 10, 20, 30, 50, and 100. Among them, 20 sized token subsequences give us better results in terms of top-k accuracy and MRR. So, we have extracted subsequences of 20 tokens from the testing dataset (developed in previous study), and moved the sliding window one step ahead to obtain further subsequences.  Out of a total of 28,197 subsequences containing 20 tokens each, we selected 735 subsequences containing the \big \langle soc\big \rangle~token, which indicates the start of a clone method. These subsequences are treated as input queries to generate DeepClone output. We passed these subsequences one by one to our DeepClone model, and we keep on predicting new tokens with nucleus sampling (threshold value 0.95) until the meta-token \big \langle eoc\big \rangle~(i.e.~end of clone) appears. We use text generation script\footnote{https://github.com/huggingface/transformers/blob/master/examples/text-generation/run\_generation.py} of HuggingFace Transformer Library in this case. Note that certain parameters, such as the number of subsequences and size of tokens in each subsequence are chosen to perform a preliminary evaluation, which can be fine-tuned and optimized in a follow-up study. The focus of the paper here is to demonstrate the feasibility of our methodology for recommending real clones.

\subsection{Retrieving Code Clones from the Search Corpus}
The output from the previous step contains set of tokens of context along with the predicted tokens up till the \big \langle eoc\big \rangle~token. In this step, we extract only those tokens, which are between \big \langle soc\big \rangle~and \big \langle eoc\big \rangle~tokens (inclusive) from the DeepClone output (see DeepClone output step in Example Table~\ref{tab:example1} and Table~\ref{tab:example2}). We apply an IR technique to retrieve top-10 similar results with the generated DeepClone output from the search corpus. IR techniques, in general, are used to discover the significant documents in a large collection of documents, which match a user’s query. Their main goal is to identify the significant information that satisfies the user information needs. An IR-based code retrieval method in particular usually extracts from a query a set of keywords and then search for the keywords in code repositories \cite{nie2016query}. 
% lu2015query,lv2015codehow, vinayakarao2017anne
% The basic idea behind embedding is to use vectors of various dimensions to represent entities numerically, which makes it easier for computers to understand them for various downstream tasks. Most of the algorithms in machine learning cannot process strings or plain text in their raw form. Instead, they require numbers as inputs to be able to function. By transforming words into vectors, word embeddings therefore allows us to process the huge amount of text data and make them fit for machine learning algorithms. They are mapped into a vector space. The idea behind it is that words with similar context occupy close spatial positions. In other words, words used in a similar context are close to each other while words that aren’t are far away from each other in the vector space. cambridge2009introduction rigby2013discovering

We have used TF-IDF word embedding based IR technique for retrieving the most similar real clone methods on the basis of DeepClone output. TF-IDF (Term Frequency-Inverse Document Frequency \cite{dillon1983introduction}) is often used in IR and text mining. A survey conducted in 2015 showed that 70\% of text-based recommendation systems in digital libraries use TF–IDF \cite{beel2016paper}. Similarly, in the past many researchers have applied TF-IDF technique to retrieve code elements  \cite{kim2018facoy,luan2019aroma}. TF-IDF is a weighting scheme that assigns each term in a document a weight based on its term frequency (TF) and inverse document frequency (IDF). In our context, TF-IDF is looking at the term overlap, i.e. the number of shared tokens between the two clone methods in question (and also how important/significant those tokens are in the clone methods). We use TF-IDF with unigrams as terms to transform clone methods into numeric vectors, that can easily be compared by quickly calculating cosine similarities. If a term appears frequently in a clone method, that term is probably important in that method: term frequency is simply the number of times that a term appears in a method. However, if a term appears frequently in many clone methods, that term is probably less important generally.  Inverse-document frequency is the logarithmically-scaled fraction of clone methods in the corpus in which the term appears. The terms with higher weight scores (high tf \emph{and} idf) are considered to be more important. We first transform clone methods existing in the search corpus and DeepClone output into TF-IDF vectors using equation ~\ref{eq:eq1}. 

\begin{equation}
\label{eq:eq1}
TF-IDF(i, j) = (1 + \log (TF(i, j)) . \log( \frac{J}{DF(i)})
\end{equation}

where \textit{TF (i, j)} is the count of occurrences of feature \textit{i} in clone method \textit{j}, and \textit{DF (i)} is the number of clone methods in which feature \textit{i} exists. \textit{J} is the total number of clone methods. During retrieval, we create a normalized TF-IDF sparse vector from the DeepClone output as query, and then take its dot product with the feature matrix. Since all vectors are normalized, the result contains the cosine similarity between the feature vectors of the query and of every clone method. We then return the list of clone methods ranked by their cosine similarities. 

% \usepackage{multirow}
\begin{table*}
\caption{An example containing scenarios such as exact match, functionality type match and method name mentioned in the context\label{tab:example1}}
\footnotesize{
\begin{tabular}{|l|l|l|l|l|l|l|l|l|l|l|l|}
\hline
\textbf{Context}                                                                                           & \multicolumn{11}{l|}{
\begin{minipage}{0.6\textwidth}`
\vspace{0.1mm}
% \centering
\includegraphics[width=0.6\textwidth]{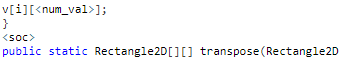}
\vspace{0.1mm}
\end{minipage}
}                                                                                                                                              \\ \hline
\multirow{2}{*}{\textbf{Ground truth (GT)}}                                                              & \multicolumn{11}{l|}{\begin{minipage}{0.7\textwidth}
\vspace{0.1mm}
% \centering
\includegraphics[width=0.7\textwidth]{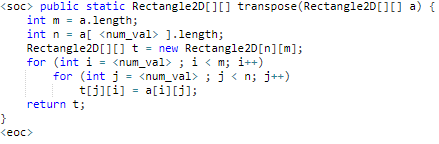}
\vspace{0.1mm}
\end{minipage}}                                                                                                                                              \\ \cline{2-12} 
                                                                                                         & \textbf{Perplexity} &  \multicolumn{10}{l|}{1.541}                   
\\ \hline
% \multirow{2}{*}{\textbf{\begin{tabular}[c]{@{}l@{}}DeepClone output \\ (DCO)\end{tabular}}} & \multicolumn{11}{l|}{\begin{minipage}{0.7\textwidth}
% \vspace{0.1mm}
% % \centering
% \includegraphics[width=0.7\textwidth]{images/exp4-topk/transPO.PNG}
% \vspace{0.1mm}
% \end{minipage}}                                                                                                                                              \\ \cline{2-11} 
%                                                                                                          & \textbf{Perplexity}: 3.54 & \textbf{DCO vs GT} & \multicolumn{8}{l|}{\begin{tabular}[c]{@{}l@{}}\textbf{ROUGE-1: } {[}P: 0.816,  R:0.857, F: 0.836{]},  \textbf{ROUGE-2: } {[}P: 0.676, R: 0.711, F: 0.693{]}, \\ \textbf{ROUGE-L: } {[}P: 0.862, R: 0.893, F:  0.877{]}\end{tabular}} \\ \hline                                       

\multirow{3}{*}{\textbf{DeepClone output (DCO)}}                                                                          & \multicolumn{11}{l|}{\begin{minipage}{0.7\textwidth}
\vspace{0.1mm}
% \centering
\includegraphics[width=0.7\textwidth]{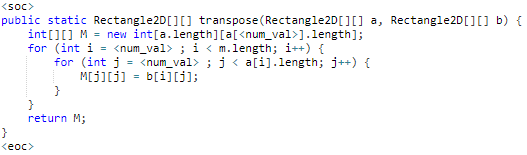}
\vspace{0.1mm}
\end{minipage}}                                                                                                              \\ \cline{2-12}

& \textbf{Perplexity}: 3.54  & \textbf{DCO vs GT} & \multicolumn{9}{l|}{\begin{tabular}[c]{@{}l@{}}\textbf{ROUGE-1: } {[}P: 0.816,  R:0.857, F: 0.836{]}, \textbf{ROUGE-2: } {[}P: 0.676, R: 0.711, F: 0.693{]}, \\ \textbf{ROUGE-L: } {[}P: 0.862, R: 0.893, F: 0.877{]}\end{tabular}} \\ \cline{2-12} 
                                                                                                                                 \hline

\multirow{3}{*}{\textbf{Top 1}}                                                                          & \multicolumn{11}{l|}{\begin{minipage}{0.7\textwidth}
\vspace{0.1mm}
% \centering
\includegraphics[width=0.7\textwidth]{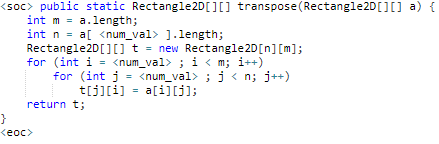}
\vspace{0.1mm}
\end{minipage}}                                                                                                              \\ \cline{2-12}

& \textbf{Perplexity}: 1.541 & \textbf{Top 1 vs DCO} & \multicolumn{9}{l|}{\begin{tabular}[c]{@{}l@{}}\textbf{ROUGE-1: } {[}P: 0.816, R: 0.857, F: 0.836 {]},  \textbf{ROUGE-2: } {[}P: 0.676, R: 0.711, F: 0.693{]}, \\ \textbf{ROUGE-L: } {[}P: 0.862, R: 0.893, F: 0.877{]}\end{tabular}} \\ \cline{2-12} 
                                                                                                                                 \hline
\multirow{3}{*}{\textbf{Top 2}}                                                                          & \multicolumn{11}{l|}{\begin{minipage}{0.7\textwidth}
\vspace{0.1mm}
% \centering
\includegraphics[width=0.7\textwidth]{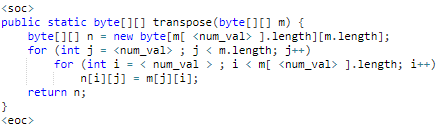}
\vspace{0.1mm}
\end{minipage}}                                                                                                                                              \\ \cline{2-12}                                                                                            & \textbf{Perplexity}:  1.63   & \textbf{Top 2 vs DCO} & \multicolumn{9}{l|}{\begin{tabular}[c]{@{}l@{}}\textbf{ROUGE-1} {[}P: 0.777, R: 0.86, F: 0.816{]}, \textbf{ROUGE-2: } {[}P: 0.627, R: 0.696, F: 0.66{]}, \\ \textbf{ROUGE-L} {[}P: 0.828, R: 0.923, F: 0.873{]}\end{tabular}} \\ \cline{2-12} 
                                                                                                        \hline
\end{tabular}
}
\end{table*}

\begin{table*}
\caption{An example containing scenarios such as functionality type match and method name not mentioned in the context\label{tab:example2}}
\footnotesize{
\begin{tabular}{|l|l|l|l|l|l|l|l|l|l|l|l|}
\hline
\textbf{Context}                                                                                           & \multicolumn{11}{l|}{
\begin{minipage}{0.55\textwidth}
\vspace{0.2mm}
% \centering
\includegraphics[width=0.55\textwidth]{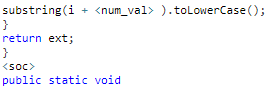}
\vspace{0.1mm}
\end{minipage}
}                                                                                                                                              \\ \hline
\multirow{2}{*}{\textbf{Ground truth (GT)}}                                                              & \multicolumn{11}{l|}{\begin{minipage}{0.7\textwidth}
\vspace{0.1mm}
% \centering
\includegraphics[width=0.7\textwidth]{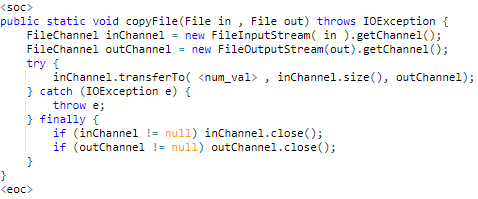}
\vspace{0.1mm}
\end{minipage}}                                                                                                                                              \\ \cline{2-12} 
                                                                                                   & \textbf{Perplexity} &  \multicolumn{10}{l|}{1.21}
\\ \hline
\multirow{2}{*}{\textbf{\begin{tabular}[c]{@{}l@{}}DeepClone output \\ (DCO)\end{tabular}}} & \multicolumn{11}{l|}{\begin{minipage}{0.75\textwidth}
\vspace{0.2mm}
% \centering
\includegraphics[width=0.75\textwidth]{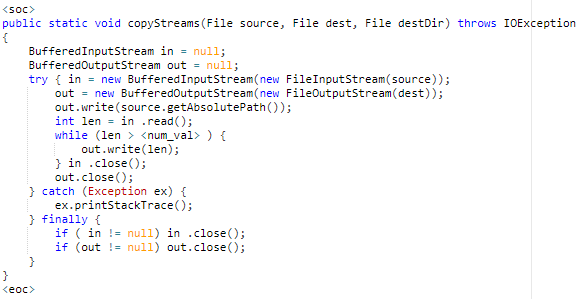}
\vspace{0.1mm}
\end{minipage}}                                                                                                                                              \\ \cline{2-12} 
                                                                                                         & \textbf{Perplexity} & 9.246
                                                          & \textbf{DCO vs GT} & \multicolumn{8}{l|}{\begin{tabular}[c]{@{}l@{}}\textbf{ROUGE-1: } {[}P:0.511,  R: 0.728, F: 0.601{]}, \textbf{ROUGE-2: } {[}P:0.331, R:0.473, F: 0.389{]}, \\ \textbf{ROUGE-L: } {[}P:0.543, R:0.676, F: 0.602{]}\end{tabular}} \\ \hline                                               
                                                                                                         
\multirow{3}{*}{\textbf{Top 1}}                                                                          & \multicolumn{11}{l|}{\begin{minipage}{0.7\textwidth}
\vspace{0.1mm}
% \centering
\includegraphics[width=0.7\textwidth]{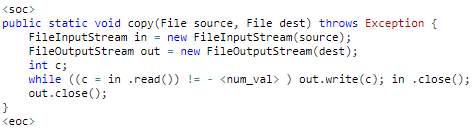}
\vspace{0.1mm}
\end{minipage}}                                                                                                              \\ \cline{2-12} 
& \textbf{Perplexity}: 1.439  & \textbf{Top 1 vs DCO} & \multicolumn{9}{l|}{\begin{tabular}[c]{@{}l@{}}\textbf{ROUGE-1: } {[}P: 0.443, R: 0.853, F: 0.583{]}, \textbf{ROUGE-2: } {[}P: 0.331, R: 0.642, F: 0.437{]}, \\ \textbf{ROUGE-L: } {[}P: 0.565, R: 0.812,F: 0.667{]}\end{tabular}} \\ \cline{2-12} 
                                                                                                        \hline
\multirow{3}{*}{\textbf{Top 2}}                                                                          & \multicolumn{11}{l|}{\begin{minipage}{0.75\textwidth}
\vspace{0.1mm}
% \centering
\includegraphics[width=0.75\textwidth]{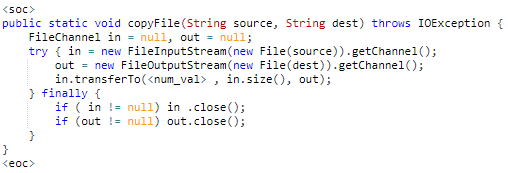}
\vspace{0.1mm}
\end{minipage}}                                                                                                                                              \\ \cline{2-12} 
                                                                                                         & \textbf{Perplexity}: 1.321  & \textbf{Top 2 vs DCO} & \multicolumn{9}{l|}{\begin{tabular}[c]{@{}l@{}}\textbf{ROUGE-1} {[}P:0.618, R: 0.835 ,F: 0.711{]}, \textbf{ROUGE-2} {[}P:0.454, R: 0.615, F: 0.522{]},  \\ \textbf{ROUGE-L: } {[}P: 0.522, R: 0.686, F: 0.593{]}\end{tabular}} \\ \cline{2-12} 
                                                                                                         \hline
\end{tabular}
}
\end{table*}
% Cosine similarity is a very common method of content-based evaluation. , there exist various methodologies \cite{gali2016similarity} and each one has a different way of calculating similarities to achieve specific goals.
\section{Empirical Evaluation}
To measure the naturalness of different clone methods, we use the perplexity score, while the quality of DeepClone output in terms of ground truth (GT) and top-k recommended clone methods is measured with ROUGE \cite{lin2004rouge} score, which is a measure to compare machine generated text/code with human written text/code. Furthermore, we evaluate whether top-k recommended clone methods match with the ground truth or not. For this purpose, we calculate top-k accuracy and MRR for exact match and functionality type match with the ground truth. All these measures show that the proposed approach significantly helps in generating real clone methods matching the user context.

\begin{table}
\label{tbl:dcovstopk}
\caption{Empirical evaluation results between DeepClone output (DCO) and top 10 recommended clones}
\footnotesize{
\begin{tabular}{|l|l|l|l|}
\hline
                    & \textbf{Top 1} & \textbf{Top (2-4)} & \textbf{Top (5-10)} \\ \hline
\textbf{ROUGE1}     & \multicolumn{3}{l|}{\textbf{}}                            \\ \hline
\textbf{Precision}  & 0.666 $\pm$  0.204        & 0.637 $\pm$ 0.197             & 0.612 $\pm$ 0.196             \\ \hline
\textbf{Recall}     & 0.599 $\pm$ 0.214        & 0.587 $\pm$  0.181           & 0.576 $\pm$ 0.181             \\ \hline
\textbf{F-measure} & 0.552 $\pm$  0.157       & 0.559 $\pm$   0.189           & 0.567    $\pm$  0.182         \\ \hline
\textbf{ROUGE-2}    & \multicolumn{3}{l|}{}                                     \\ \hline
\textbf{Precision}  & 0.485     $\pm$ 0.225    & 0.457    $\pm$ 0.21         & 0.425  $\pm$ 0.203            \\ \hline
\textbf{Recall}     & 0.403 $\pm$  0.187       & 0.387    $\pm$ 0.168         & 0.374   $\pm$   0.167         \\ \hline
\textbf{F-measure} & 0.362 $\pm$ 0.149        & 0.353    $\pm$ 0.173         & 0.361 $\pm$     0.163         \\ \hline
\textbf{ROUGE-L}    & \multicolumn{3}{l|}{}                                     \\ \hline
\textbf{Precision}  & 0.631 $\pm$  0.175         & 0.625 $\pm$    0.17         & 0.601 $\pm$    0.167           \\ \hline
\textbf{Recall}     & 0.591 $\pm$ 0.177        & 0.58     $\pm$ 0.154         & 0.565 $\pm$     0.148         \\ \hline
\textbf{F-measure} & 0.556 $\pm$  0.136       & 0.549    $\pm$ 0.145         & 0.548 $\pm$ 0.151             \\ \hline
\end{tabular}
}
\end{table}

 \begin{table*}
\footnotesize{
\parbox{.33\linewidth}{
\centering
\label{tbl:results_gtpo}
\caption{Empirical evaluation results between DeepClone output (DCO) and ground truth (GT)}
\begin{tabular}{|l|l|}
\hline
                   & \textbf{DCO vs GT} \\ \hline
% \textbf{Cosine}    & 0.675 $\pm$ 0.145          \\ \hline
\textbf{ROUGE-1}   &                   \\ \hline
\textbf{Precision} & 0.667 $\pm$ 0.192           \\ \hline
\textbf{Recall}    & 0.559   $\pm$ 0.226      \\ \hline
\textbf{F-measure}  & 0.56 $\pm$     0.185    \\ \hline
\textbf{ROUGE-2}   &                   \\ \hline
\textbf{Precision} & 0.479 $\pm$    0.217     \\ \hline
\textbf{Recall}    & 0.398 $\pm$ 0.218         \\ \hline
\textbf{F-measure}  & 0.4 $\pm$  0.202       \\ \hline
\textbf{ROUGE-L}   &                   \\ \hline
\textbf{Precision} & 0.652  $\pm$  0.165      \\ \hline
\textbf{Recall}    & 0.586  $\pm$  0.183      \\ \hline
\textbf{F-measure}  & 0.599  $\pm$  0.153      \\ \hline
\end{tabular}
}
%%%%TODO display standard deviation symbol
\hfill
\parbox{.23\linewidth}{
\centering
\caption{Perplexities}
\label{tbl:perplexities}
\begin{tabular}{|l|l|}
\hline
            & \textbf{Perplexity} \\ \hline
\textbf{DCO} & 11.624 $\pm$ 5.892          \\ \hline
\textbf{GT} & 2.047 $\pm$ 0.848          \\ \hline
\textbf{1}  & 1.79 $\pm$ 0.716          \\ \hline
\textbf{2}  & 1.851   $\pm$ 0.726        \\ \hline
\textbf{3}  & 1.795 $\pm$  0.594         \\ \hline
\textbf{4}  & 1.800 $\pm$  0.491         \\ \hline
\textbf{5}  & 1.874 $\pm$  0.544         \\ \hline
\textbf{6}  & 1.907 $\pm$  0.612         \\ \hline
\textbf{7}  & 1.887 $\pm$  0.578         \\ \hline
\textbf{8}  & 1.863  $\pm$ 0.505         \\ \hline
\textbf{9}  & 1.855 $\pm$  0.545         \\ \hline
\textbf{10} & 1.859 $\pm$  0.578         \\ \hline
\end{tabular}

}
\hfill
\parbox{.40\linewidth}{
\caption{MRR and Top-k accuracies}
\label{tab:exactmatches}
\centering
\begin{tabular}{|l|l|l|}
\hline
\textbf{}       & \textbf{Exact Match} & \textbf{Functionality Type Match} \\ \hline
\textbf{MRR}    &    0.283                   &       0.738                         \\ \hline
\textbf{Top-1}  &     0.233                   &     0.692                         \\ \hline
\textbf{Top-3}  &   0.316                   &       0.770                       \\ \hline
\textbf{Top-5}  &   0.355                   &       0.797                       \\ \hline
\textbf{Top-10} &   0.393                   &       0.841                       \\ \hline
\end{tabular}
}
}
\end{table*}

% \begin{tabular}{|l|l|l|l|l|l|}
% \hline
%                                                                   & \textbf{MRR} & \textbf{Top-1} & \textbf{Top-3} & \textbf{Top-5} & \textbf{Top-10} \\ \hline
% Exact Match                                                        & 0.283        & 0.233          & 0.316          & 0.355          & 0.393           \\ \hline
% \begin{tabular}[c]{@{}l@{}}Functionality Type\\ Match\end{tabular} &  0.738             &    0.692             &  0.770              &   0.797              &   0.841              \\ \hline
% \end{tabular}

\subsection{Perplexity}
% There are various studies has been conducted on finding the buggy code \cite{ray2016naturalness,karampatsis2020big}. mikolov2011empirical
In previous work, it has been observed that n-gram language models can detect defects as they are less “natural” than correct code \cite{ray2016naturalness}. Similarly, Karampatsis et al. \cite{karampatsis2020big} have noted that defective lines of code have a higher cross-entropy ($\sim$perplexity, to be explained later in this section) than their correct(ed) counterparts. Using the original DeepClone model, the predicted clone method is a buggy snippet, as we observe and can be expected from probabilistic language models. We measure and argue about perplexity scores for the original DeepClone output versus the real clone methods around these angles of naturalness and potential bug density. We expect original DeepClone output to have relatively high perplexity. Perplexity is used to measure the degree of accurately predicting sample data using a language model. At each point in a sequence, it gives an estimate of the average number of code tokens to select from \cite{allamanis2013mining}. Perplexity represents a probability distribution over a subsequence or even an entire dataset, and is widely used as a natural evaluation metric for language models. The formula for perplexity is presented in Equation~\ref{eq:eq6}:

\begin{equation}
\label{eq:eq6}
P(L) = exp(-\frac{1}{M} \sum_{i}^{M} \log P(t_i|t_0 : t_{i-1}))
\end{equation}

where $P(t_i | t_0:t_{i-1})$ is the conditional probability assigned by the model to the token $t$ at index $i$. By applying $log$ of conditional probability, cross-entropy loss is calculated. $M$ refers to the length of tokens. Hence, perplexity is an exponentiation of the average cross entropy loss from each token $[0, M]$. To assess how the original output of DeepClone model differs from the real clone code methods, we find the perplexity scores of the clone method predicted by DeepClone and top-k most similar retrieved clone methods. DeepClone output is potentially more noisy and less natural as compared to ground truth (GT) and top-10 recommended snippets in the samples. Table ~\ref{tbl:perplexities} depicts the mean perplexities of top-10 recommendations and DeepClone output. This clearly displays that DeepClone output has substantial \emph{noise} in it (confirming our observations), some of which attributes to a high density of errors. On the other hand, the top-10 retrieved snippets have relatively low perplexities, which indicates that they are highly natural and less noisy as compared to DeepClone output. There are slight variations in the perplexity values of Top-10 samples, which can be attributed to various factors such as the type of functionality, the number of clone method snippet trained in the DeepClone model, and inner similarity among the clone methods' type. These factors have been discussed in detail in our previous work \cite{hammad2020deepclone}.

\subsection{Exact Match Evaluation}
We collect the top 10 recommended clone methods retrieved by our approach, and compute the top-k accuracy (the fraction of times the ground truth clone method appears in the top k recommended clone methods) for k $\in$ [1, 10]. Moreover, we measure the Mean Reciprocal Rank (MRR) scores for the recommendations. A simplified description of MRR is that it averages top-k accuracy across various k. In this specific scenario k $\in$ [1, 10] since the methodology output a list of top-10 recommended clone methods. The MRR is a rank-based evaluation metric, which produces a value between 0 and 1, where a value closer to 1 indicates a better clone method recommendation system. The reciprocal rank of a query response is the multiplicative inverse of the rank of the first correct answer, while MRR is the average of reciprocal ranks. Table~\ref{tab:exactmatches} shows the top-k accuracy as well as the MRR score. The results display that our methodology has the capability of identifying an exact match between ground truth and top-k recommend clone methods. Table~\ref{tab:example1} displays that top-1 recommended clone method exactly matches the ground truth.

%  We calculate the total number of exact matches between the functionality type ids of recommended clone method and ground truth (GT) for each input query (Table ~\ref{tbl:functionalityType}). Moreover,
\subsection{Functionality Type Evaluation}
BigCloneBench contains references of multiple implementations (i.e.~clones) of specific functionalities. It contains validated clone methods belonging to 43 different functionalities, for instance, "copy file" functionality contains 3055 different implementations. Further details can be found from our previous paper\cite{hammad2020deepclone}, and BigCloneBench dataset\cite{hammad2020deepclone}. Hence, it is possible to have recommended clone methods that do not exactly match the ground truth but match its functionality. For instance, Table~\ref{tab:example2} displays top-1 and top-2 clone methods belonging to the same functionality type as the ground truth (GT). So, both implementations can potentially satisfy the user's need. For this purpose, we extract the functionality id of the ground truth and recommended list of top-k clone methods against each context. We calculate top-k accuracy and MRR accordingly (see Table~\ref{tab:exactmatches}). The results indicate that we can recommend methods with the matching functionality (if not the exact implementation) with high accuracy, summarized with an MRR score of 0.738. 

% the fraction of times the functionality id of ground truth appears in the functionality id of top k recommended clone methods (top-k accuracy). Finally, we calculate means reciprocal rank (MRR) which is the. 

% The standard evaluation measure in summarization is ROGUE, which quantifies lexical overlaps between a system generated and a reference summary as a score for evaluating quality. It has been used to evaluate automatic summarization of texts as well as machine translation. to the reference clone methods in the annotated BigCloneBench dataset nallapati2016abstractive
\subsection{ROUGE Metrics}\label{sec:rouge}
We have also evaluated our approach by using ROUGE (Recall-Oriented Understudy for Gisting Evaluation) \cite{lin2004rouge}. It is designed to compare an automatically generated summary or translation against a set of reference summaries (typically human-generated). The ROUGE score (scores) allows us to measure the quality of text summarization by computing the frequency of overlapping n-grams between the produced summary and the reference one(s). Various forms of overlapping units, such as n-grams, word sequences, and word pairs, are counted between the auto-generated summary and the reference summaries. In our context, it helps us to automatically determine the quality of original DeepClone output by comparing it with the ground truth and top-10 recommended clone methods. ROUGE doesn't try to assess how fluent the clone method is. It only tries to assess the adequacy, by simply counting how many n-grams in the DeepClone output matches the n-grams in the ground truth and top-k recommended clone methods. Because ROUGE is based only on token overlap, it can determine if the same general concepts are discussed between an automatic summary and a reference summary, but it cannot determine if the result is coherent or the clone method is semantically correct. 
%flow together in a sensible manner.
High-order n-gram ROUGE measures try to judge fluency to some degree. In this paper, we evaluate the quality of DeepClone output with respect to the ground truth and recommended top-10 clone methods by calculating the scores of ROUGE-1, ROUGE-2, and ROUGE-L as most authors use them for automatic evaluation score besides human evaluation \cite{moeed2020evaluation,paulus2017deep}. We have calculated precision (P), recall (R), and F-measure (F) of ROUGE-1 ROUGE-2, and ROUGE-L between various combinations such as DeepClone output and ground truth (DCO vs GT), DeepClone output and top-10 recommended clone methods (top-1...10 vs DCO). ROUGE-1 refers to the overlap of unigrams between some reference output and the output to be evaluated. ROUGE-2, in turn, checks for bigrams instead of unigrams. The reason one would use ROUGE-1 over or in conjunction with ROUGE-2 (or other finer granularity ROUGE measures), is to also indicate fluency as a part of the evaluation. The intuition is that the prediction is more fluent if it more closely follows the word orderings of the reference snippet. Finally, ROUGE-L measures longest matching sequence of tokens between machine generated text/code and human produced one by using longest common subsequence (LCS). Using LCS has a distinguishing advantage in evaluation: it captures in-sequence (i.e.~sentence level flow and word order) matches rather than strict consecutive matches. 

DeepClone predicted clone method can be extremely long, capturing all tokens in the retrieved clone methods, but many of these tokens may be useless, making it unnecessarily verbose. This is where precision comes into play. It measures what portion of the DeepClone output was in fact relevant and desirable to be kept with respect to the reference output.

\begin{equation}
\label{eq:eqprecision}
\text{Precision}=\frac{\text{\# of overlapping tokens}}{\text{total tokens in the predicted output}}
\end{equation}

Recall in the context of ROUGE measures what portion of the reference output was successfully captured by the DeepClone output.
\begin{equation}
\label{eq:eqrecall}
\text{Recall}=\frac{\text{\# of overlapping tokens}}{\text{total \# tokens in reference snippet}}
\end{equation}

We also report the F-measure which provides a single score that balances both the concerns of precision and recall.

\begin{equation}
\label{eq:eqpfscore}\text{F-Measure} = 2*\frac{\text{Precision * Recall}}{\text{Precision + Recall}}
\end{equation}

% Mean cosine similarity between predicted output (PO) from DeepClone model and ground truth (GT) is 0.675. Mean cosine similarity between top-1 result and ground truth is almost 0.688, which depicts how much our information retrieval technique is good in retrieving real clone methods. Similarly, mean cosine similarity between top-1 result and DeepClone output is almost 0.697, which is considered to have reasonable cosine similarity. We have measured various similarity measures to evaluate our strategy.  ROUGE score is also reasonable.

\section{Results and Discussion}
\label{sec:results}
%Properly introduce the results, X in table, summarize them, what do they show, refer to and reflect on examples where applicable etc.We observe high perplexity scores for DeepClone output (11.624) as compared to ground truth (2.047) and top-10 recommended clone methods (range from 1.79 till 1.907). 
The approach we have proposed leads to promising results.  We observe quite high mean perplexity scores and standard deviation of DeepClone output (11.624 $\pm$ 5.892), This indicates high noise and less natural code, which is a known problem of neural language generation \cite{li2017adversarial,shao2017generating}. However, we notice quite low mean perplexity scores and standard deviation for the ground truth (2.047 $\pm$ 0.848) and top-10 recommended clone methods (range from 1.79 $\pm$ 0.716 till 1.907 $\pm$ 0.612) against the set of 735 input queries. These numbers show that our methodology can improve the original prediction by DeepClone, resulting in more natural snippets. We have further calculated top-k accuracy and MRR involving an exact match of the recommended clone methods with the ground truth. We achieve an accuracy of 39.3\% in the top 10 recommended clone methods and MRR as 28.3\% (see Table ~\ref{tab:exactmatches}). In a fair share of the cases, we can find exactly the same clone method as in the ground truth. As BigCloneBench contains references to various implementations for each of the 43 functionalities, it is quite possible that the user is recommended a different snippet than the ground truth, yet implementing the same functionality. This alternative can also help the developer achieve their goal. To assess such cases, we have calculated the top-k accuracy and MRR taking alternative implementations into account. We achieve quite a high accuracy, notably 84.1\% in the top 10 recommended clone methods, as well as 73.8\% MRR in terms of functionality type match with the ground truth  (see Table ~\ref{tab:exactmatches}). This is a major improvement over the exact match scores and further reinforces the claims we make for our approach. 

Furthermore, we have measured different ROUGE scores, i.e.~ROUGE-1, ROUGE-2, and ROUGE-L, to evaluate the similarity (and the quality to a certain extent) of the DeepClone output to ground truth and top-10 recommended clone methods. After retrieving the top-10 recommendations, we have calculated the precision, recall and F-measure (as explained in Section~\ref{sec:rouge}) for each ROUGE metric. Note that our approach ranks the recommendations with respect to their cosine similarity. Thus it is possible that when we compare DeepClone output with top-10 recommended clone methods using ROUGE, precision, recall and F-measure may not come in a strict descending order given the different characteristics of each ROUGE metric (see Table ~\ref{tbl:dcovstopk}). Cosine similarity is a very common method for content-based evaluation, while ROUGE scores are used to evaluate machine generated against human produced content \cite{lin2004rouge}. That is why we observe that F-measure for ROUGE-1 of Top (2-4) is slightly higher than Top-1 \ref{tbl:dcovstopk}. 

In our qualitative investigation, we experienced two different scenarios based on the recommended output. The first one is the ideal scenario when one of the top-k recommended clone methods exactly match the ground truth. In the example given in Table~\ref{tab:example1}, "transpose" clone method implementation at top-1 exactly matches the ground truth. This scenario gives the best results. The second scenario is when none of the top-k recommended methods exactly match the ground truth but at least one of the top-k recommended clone method functionality type matches with functionality type of the ground truth. In Table~\ref{tab:example2}, although top-1 and top-2 recommended clone methods do not exactly match with the ground truth, they belong to the same functionality type "copy file". The main advantage of our methodology is that even if the recommended clone methods do not exactly match the ground truth, still they would be usually implementing the same functionality as the ground truth method, and might satisfy the user's need. 

Similarly, there are two scenarios based on the input context. The first scenario is when the context contain the name of method. It is straightforward for the neural language technique to generate the predicted clone method following the given method name and current context. Table ~\ref{tab:example1} gives an example of this scenario, where "transpose" method name is mentioned in the context and our approach recommends clone methods as top-1 and top-2, whose functionality type matches with the functionality type of the ground truth. The second scenario is based on the context that does not contain a method name. This can have two different output sub-scenarios. The first one is when the functionality type of the recommended clone method and the ground truth do not match. As we see in Table~\ref{tab:example3}, the context does not have the full signature of the clone method. This makes the generated output by DeepClone using nucleus sampling deviate from the functionality type of the ground truth. Ground truth belongs to "copy file" functionality, while DeepClone output belongs to "delete directory" functionality type, which eventually leads to TF-IDF recommending clone methods as top-1 and top-2 on the basis of "delete directory". Such scenarios eventually result in low and largely deviating ROUGE scores between the DeepClone output and the ground truth (see Table ~\ref{tbl:results_gtpo} and the example in Table ~\ref{tab:example3}). There we observe that it also affects the other evaluation measures involving exact and functionality type matches. These clone methods eventually may not fulfil the desired goal of the user (see Table ~\ref{tab:exactmatches}). So, it might be useful to guide the users to include the clone method name in the context for better results. The other output sub-scenario is when we manage to successfully generate DeepClone output whose functionality type matches with the ground truth. In Table~\ref{tab:example2}, "copy file" method name is not mentioned in the context, but the functionality type of the DeepClone output matches with the ground truth, which eventually helps TF-IDF to retrieve clone methods on the basis of DeepClone output. We notice that the total number of "copy file" clone methods used in DeepClone training are 2,454, which allows nucleus sampling to generate DeepClone output closer to ground truth in example ~\ref{tab:example2}. Overall we believe our approach yields very promising results and can assist the developers by recommending real and accurate clone methods. 

\section{Limitations and Threats to Validity}
The proposed approach is the first step towards recommending real code clone methods on the basis of user context. However, it has certain limitations as well. Despite the fact that the dataset used in this study is collected from a well-known clone code dataset (BigCloneBench), it does not necessarily mean the codebase used in this study represents Java language source code entirely (a threat to external validity in terms of generalizability). Another issue is that we have selected and used various parameter/threshold values and techniques with the goal of showcasing the feasibility of our approach. As an example, for generating predicted clone methods, we only used nucleus sampling with threshold value of 0.95 \cite{holtzman2019curious}. There are various other text generation methods such as beam search \cite{vijayakumar2018diverse}, sampling with temperature \cite{ackley1985learning}, and top-k sampling \cite{fan2018hierarchical}, which can be explored for generating clone methods on the basis of user context. Similarly, threshold values can be tuned to get the best results. Based on experimentation, we determined certain parameters (e.g.~735 as the number of subsequences, 20 as the number of tokens per subsequence) aiming to demonstrate a preliminary evaluation of our methodology. However, by having different parameters, e.g.~ having subsequences of different sized tokens, and using the complete set of queries, we can have different results. 

Another limitation involves the normalization step we have performed. We have replaced integer, float, binary, and hexadecimal constant values with the \big \langle num\_val\big \rangle~meta-token. Similarly, we have replaced string and character values with \big \langle str\_val\big \rangle. This reduces our vocabulary size, which leads to faster training of the model, but also reduces the vocabulary of the predictions. We nevertheless note that technique has been used by several researchers in the same manner for data preparation \cite{white2015toward,karampatsis2020big}. Similarly, in order to have fair comparison between DeepClone output and clone methods available in search corpus, we have built a search corpus in the same format as we have used for DeepClone. This helps the TF-IDF technique to recommend clone methods accordingly. %In the future, we plan to replace these meta tokens with original constant values of real clone methods, so that these clone methods can work in Java based IDE tools such as Eclipse, without leading to syntax errors. Moreover, \big \langle soc\big\rangle~and \big \langle eoc\big\rangle~tokens help nucleus sampling to generate DeepClone output. Same meta tokens have been used in search corpus to help TF-IDF to have fair enough recommended clone methods. In the future, we aim to remove these meta-tokens, so that these recommended clone methods can work directly in IDEs. 

In this study, we only apply TF-IDF, an IR technique to retrieve the most similar real clone methods, on the basis of the predicted clone method. However, there are other IR techniques such as GLOVE\cite{pennington2014glove}, and word2vec\cite{mikolov2013efficient}, which can be additionally explored. We leave it for future work to comparatively assess and optimize the parameters and techniques for our approach.

\section{Related Work}
In this section, we present related work covering neural language generation techniques, and the role of recommendation systems in the field of code clones.

\subsection{Neural Language Generation}
To the best of our knowledge, no technique has been presented to improve the prediction of clone methods. However, many techniques have been introduced to improve the quality of generated program code and text. Hashimoto et al. \cite{hashimoto2018retrieve} proposed an approach to predict python code tokens, first by retrieving a training example based on the input (e.g., natural language description) and then editing it to the desired output (e.g., code). Song et al.\cite{song2016two} proposed a novel ensemble of retrieval-based and generation-based dialog systems. They first obtained a candidate response on the basis of user utterance or query by applying IR technique from a large database. Then, they passed retrieved candidate and query to an RNN-based response generator, so that the neural model is aware of more information. The generated response is then fed back as a new candidate for post-reranking. Zhou el al. have proposed Lancer \cite{zhou2019lancer}, a new context-aware code-to-code recommending tool leveraging a Library-Sensitive Language Model and a BERT model to recommend relevant code samples in real-time, by automatically analyzing the intention of the incomplete code. Lancer uses BERT model to complete an incomplete code sample, then it retrieves the relevant real code samples on the basis of Elastic search and rank them according to the deep semantic ranking scheme. The major difference with our methodology is that they used BERT model, whose intention is to complete the missing tokens in the incomplete code, while we are using fine-tuned GPT2 model in DeepClone, which is used to predict next tokens on the basis of context. DeepClone is fine-tuned on both non-clone and clone code patterns, but Lancer only used clone methods for training. Moreover, we can also generate correct DeepClone output when the context does not even contain a method name, a scenario that is apparently not covered by Lancer.

% the recommended clone mehods has been correctly identified on the basis of current context in which method name is not mentioned.

\subsection{Code Clone Recommendation Systems}
We could only find a single piece of work which suggests using clone methods for code recommendation\cite{abid2019recommending}. However, they use a different similarity measure based on API calls for recommending relevant clone methods. Clones are generally considered to be harmful for a software system, and mainly researchers work on techniques for avoiding and eliminating clones \cite{yoshida2019proactive,wang2014recommending,basit2015we,hammad2020systematic}. Clone refactoring recommendation systems have been developed for this purpose. For instance, Yoshida et al. \cite{yoshida2019proactive} proposed a proactive clone recommendation system for “Extract Method” refactoring, while Wang et al. \cite{wang2014recommending} introduced an approach for automatically recommending clones for refactoring using a decision tree-based classifier.

\section{Conclusion and Future Work}
In this work, we have proposed a recommendation system to suggest real code clones by using IR techniques. This system significantly improves the original prediction by DeepClone, a deep learning model we previously developed. We have have performed quantitative evaluation using a wide range of metrics, and qualitatively discussed additional scenarios, to support our claim. In the future, we plan to perform a comparative study by evaluating different IR techniques such as BERT; pretrained word embedding techniques such as word2vec\cite{mikolov2013efficient} and GLOVE \cite{pennington2014glove}; and code query formulation techniques \cite{lu2015query,nie2016query}. %Furthermore, we aim to develop and evaluate a visualization tool on top of our system to provide a user-friendly environment for assisting the developers. This also includes fully automatic formatting of the code rather than the semi-automatic approach we have taken in this paper. 

\section*{Acknowledgment}
We acknowledge the contribution of Dr. Sohaib Khan (CEO at Hazen.ai) for providing us valuable feedback on methodology and empirical evaluation parts.

\bibliographystyle{IEEEtran}
\bibliography{arxiv}
\clearpage
\appendices
% \section{Examples}
% \arabic{table}

\setcounter{table}{0}
\renewcommand{\thetable}{APPENDIX}
\begin{table*}
% \caption{An example containing scenarios such as functionality type of top-1 and top-2 recommended clone methods not matched with the ground truth (GT) and method name not mentioned in the input query\label{tab:example3}}

\caption{An example containing scenarios such as functionality type not matched and method name not mentioned in the context\label{tab:example3}}
\footnotesize{
\begin{tabular}{|l|l|l|l|l|l|l|l|l|l|l|l|}
\hline
\textbf{Context}                                                                                           & \multicolumn{11}{l|}{
\begin{minipage}{0.6\textwidth}
\vspace{0.2mm}
% \centering
\includegraphics[width=0.6\textwidth]{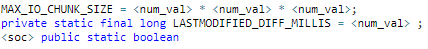}
\vspace{0.1mm}
\end{minipage}
}                                                                                                                                              \\ \hline
\multirow{2}{*}{\textbf{Ground truth (GT)}}                                                              & \multicolumn{11}{l|}{\begin{minipage}{0.8\textwidth}
\vspace{0.1mm}
% \centering
\includegraphics[width=0.8\textwidth]{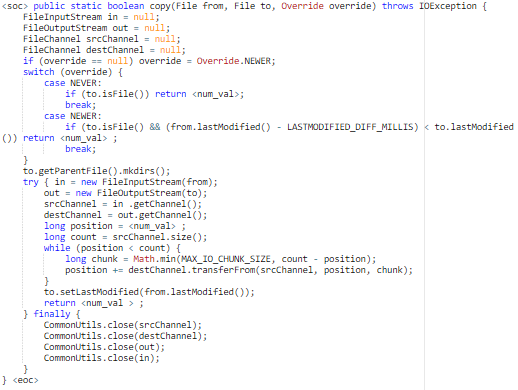}
\vspace{0.1mm}
\end{minipage}}                                                                                                                                              \\ \cline{2-12} 
                                                                                                   & \textbf{Perplexity} &  \multicolumn{10}{l|}{3.109}                             
\\ \hline
\multirow{2}{*}{\textbf{\begin{tabular}[c]{@{}l@{}}DeepClone output \\ (DCO)\end{tabular}}} & \multicolumn{11}{l|}{\begin{minipage}{0.65\textwidth}
\vspace{0.2mm}
% \centering
\includegraphics[width=0.65\textwidth]{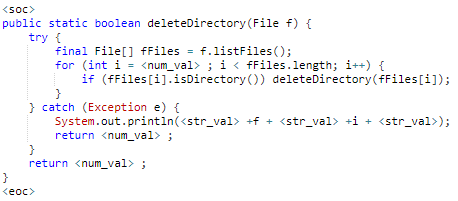}
\vspace{0.1mm}
\end{minipage}}                                                                                                                                              \\ \cline{2-12} 
                                                                                                         & \textbf{Perplexity} & 6.137
                                                          & \textbf{DCO vs GT} & \multicolumn{8}{l|}{\begin{tabular}[c]{@{}l@{}}\textbf{ROUGE-1: } {[}P:0.551,  R: 0.216, F: 0.31{]}, \textbf{ROUGE-2: } {[}P:0.295, R:0.115, F: 0.166{]}, \\ \textbf{ROUGE-L: } {[}P: 0.474, R:0.281, F: 0.353{]}\end{tabular}} \\ \hline                                               
                                                                                                         
\multirow{3}{*}{\textbf{Top 1}}                                                                          & \multicolumn{11}{l|}{\begin{minipage}{0.55\textwidth}
\vspace{0.2mm}
\includegraphics[width=0.55\textwidth]{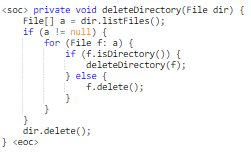}
\vspace{0.1mm}
\end{minipage}}                                                                                                              \\ \cline{2-12} 
& \textbf{Perplexity}: 1.887  & \textbf{Top 1 vs DCO} & \multicolumn{9}{l|}{\begin{tabular}[c]{@{}l@{}}\textbf{ROUGE-1: } {[}P: 0.494, R: 0.677, F: 0.571{]},  \textbf{ROUGE-2: } {[}P: 0.295, R: 0.406, F: 0.342{]}, \\ \textbf{ROUGE-L: } {[}P:  0.447, R:0.654,F: 0.531{]}\end{tabular}} \\ \cline{2-12} 
                                                                                                         \hline
\multirow{3}{*}{\textbf{Top 2}}                                                                          & \multicolumn{11}{l|}{\begin{minipage}{0.6\textwidth}
\vspace{0.2mm}
% \centering
\includegraphics[width=0.6\textwidth]{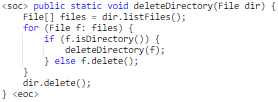}
\vspace{0.1mm}
\end{minipage}}                                                                                                                                              \\ \cline{2-12} 
                                                                                                         & \textbf{Perplexity}:1.831   & \textbf{Top 2 vs DCO} & 
                                                                                                             \multicolumn{9}{l|}{\begin{tabular}[c]{@{}l@{}}\textbf{ROUGE-1} {[}P:0.494, R: 0.786 ,F: 0.607{]},  \textbf{ROUGE-2} {[}P:0.318, R: 0.509, F: 0.392{]}, \\ \textbf{ROUGE-L: } {[}P: 0.5, R: 0.76, F: 0.603{]}\end{tabular}} \\ \cline{2-12}                                                        \hline
\end{tabular}
}
\end{table*}

\vspace{12pt}

\end{document}